\documentclass[preprint,prd,nofootinbib]{revtex4}

\usepackage{color} 
     
\usepackage{amssymb}
\usepackage{amsmath}
\usepackage{graphicx}

\begin{document}

\title{Particle production in a gravitational wave background}
\author{Preston Jones}
\email{preston.jones1@erau.edu}

\affiliation{Embry Riddle Aeronautical University, Prescott, AZ 86301}

\author{Patrick McDougall}
\email{pmcdougall@mail.fresnostate.edu}

\author{Douglas Singleton}
\email{dougs@csufresno.edu}

\affiliation{Physics Department, California State University Fresno, Fresno, CA 93740}

\date{\today}

\begin{abstract}
We study the possibility that massless particles, such as photons, are produced by a gravitational wave. That such a process should occur is implied by tree-level, 
Feynman diagrams such as two gravitons turning into two photons {\it i.e.} $g + g \rightarrow \gamma + \gamma$. Here we calculate the rate at which a 
gravitational wave creates a massless, scalar field. This is done by placing the scalar field in the background of a plane gravitational wave and calculating the 
4-current of the scalar field. Even in the vacuum limit of the scalar field it has a non-zero vacuum expectation value (similar to what occurs in the Higgs mechanism) and a non-zero current. We 
associate this with the production of scalar field quanta by the gravitational field. This effect has potential consequences for the attenuation of gravitational waves since the massless 
field is being produced at the expense of the gravitational field. This is related to the time-dependent Schwinger effect, but with the electric field replaced 
by the the gravitational wave background and the electron/positron field quanta replaced by massless scalar ``photons". Since the produced scalar quanta are massless there is no exponential suppression as occurs in the Schwinger effect due to the electron mass. 
\end{abstract}

\maketitle

\section{Introduction}

As early as 1855 Faraday recognized the possibility of a relationship between gravity and electricity \cite{Faraday1885} through his observation ``Such results, if possible, could only be exceedingly small; but, if possible, {\it i.e.} if true, no terms could exaggerate the value of the relation they would establish". More recently the potential relationship between gravity and the electromagnetic interactions has been examined for individual quanta in terms of gravitons and photons \cite{Skobelev75,Bohr14,Bohr15} using Feynman diagrams or in terms of electromagnetic waves and gravitational waves \cite{Gertsenshtein60,Jones15,Calmet16} ({\it i.e.} large collections of photons and gravitons). The perturbative, Feynman diagrammatic calculations of \cite{Skobelev75,Bohr14,Bohr15} give transitions from gravitons to photons which are consistent with Faraday's expectation that this effect is ``exceedingly small". For example in \cite{Skobelev75} it was found that the cross section for two gravitons going to two photons ($g+g \rightarrow \gamma + \gamma$) \footnote{Since here we have in mind to calculate how a gravitational plane wave, which is composed of many gravitons, is converted into a massless field, the gravitons would be taken as going in the forward direction and the massless field created from the gravitational wave would also be going in the forward direction as expected from energy-momentum conservation \cite{Modanese95}.}  is of the order $\sigma \sim 10^{-110} cm^2$ for a wave whose frequency is set by the electron rest mass $\omega \sim m_e$. For such a small cross section this process is not important even for gravitons traveling cosmological distances. The frequencies involved in the detection by LIGO of GW150914 \cite{LIGO} where much
lower than $\omega \sim m_e$, which would make the cross sections even smaller. The point of these estimates is that one gets a small,
but non-zero result for this process.. In this paper we want to examine the production of massless quanta from a gravitational wave background. This can be viewed as a gravitational variant of the Schwinger effect where a strong, static electric field can produce electron-positron pairs \cite{schwinger}. In the present case the background field is that of a gravitational wave instead of a static electric field and the particles produced are massless scalar particles (which are stand-ins for photons) instead of electrons-positrons. In the usual Schwinger effect the electron-positron production rate per unit volume is given by 
\begin{equation}
\label{e+e-}
\Gamma_{e+e-} = \frac{e^2 E_0^2}{4 \pi^3} \exp \left[\frac{-\pi m_e^2}{e E_0} \right] ~.
\end{equation}
This process is exponential suppressed by the last term in the expression above ($E_0$ is the magnitude of the electric field, $m_e$ is the electron mass and $e$ is the electron charge). In the case studied here -- gravitational field creating massless quanta -- there will be no exponential suppression since the mass associated with the fields is zero.

A final important point about taking the scalar field to be massless is that it has been shown \cite{Gibbons75} that a gravitational plane wave can not create a scalar field {\it if the scalar field is massive}. The caveat given in \cite{Gibbons75} for when it might be possible to create a scalar field from a gravitational plane wave is exactly when the scalar field is massless. This also fits in with the particle view point of reference \cite{Modanese95} where the decay of gravitons into other particles was investigated, and from very simple kinematic arguments it was shown that graviton decay was only possible when the graviton decayed into other massless particles. 

The potential significance of the process where electromagnetic radiation is produced from a gravitational wave background, is that this would lead to a weakening or attenuation of the gravitational wave, since the creation of the electromagnetic radiation would come at the expense of the gravitational wave. If the production of electromagnetic radiation via gravitational waves is significant, 
one would need to take this into account when using the detected amplitude of the gravitational wave to determine the characteristics of the event, such as the distance to the source of the gravitational waves. For example, this attenuation would mean that the source of the gravitational wave was closer than implied by the measured amplitude. Another consequence of this process is that one might
think to look for the electromagnetic radiation which was produced by the gravitational wave. In fact there is a claim \cite{loeb} that the gravitational wave detection by LIGO, GW150914, \cite{LIGO} was potentially accompanied by a $\gamma$-ray signal. Our calculations below will show that a gravitational wave might produce electromagnetic radiation, but rather than being in the $\gamma$-ray range, the
electromagnetic radiation produced would have extremely long wavelengths on the order of 100s of  kilometers. 

Previously the question of production of electromagnetic radiation from a gravitational background was examined by two of the authors \cite{Jones15} using the formalism of the Unruh-DeWitt detector. The resulting scalar field quanta production rate found in this way was small but not as small as indicated by the Feynman diagram calculations \cite{Skobelev75,Bohr14,Bohr15} for individual quanta. Based on the Unruh-Dewitt detector calculations of \cite{Jones15} it was possible that the production of electromagnetic radiation via a gravitational wave background might have an attenuating effect on the gravitational wave. This difference between the Feynman diagrammatic calculations of \cite{Skobelev75,Bohr14,Bohr15} and the Unruh-DeWitt detector calculations of \cite{Jones15} can be compared to the situation that occurs when calculating the decay rate, $\Gamma _{e+e-}$, for the Schwinger effect. The expression for $\Gamma _{e+e-}$ given in \eqref{e+e-} is  non-perturbative (this can be seen by the presence of $\exp{[-\frac{const.}{E_0}]}$) and cannot be obtained via the perturbative method of Feynman diagrams.      

In Minkowski space-time the calculation of vacuum pair production via different methods gives identical results. For example, one can calculate the Schwinger effect via the Trace-Log method originally used by Schwinger or via the ``scattering/tunneling" of some charged field by the potential due to the background, uniform electric field and the results are the same (this comparison of different methods of calculating the Schwinger effect can be found in \cite{holstein98,holstein99} as well as in \cite{Padmanabhan99}). However, in curved space-times, different methods for calculating particle production can give different results for the production rate as discussed in several papers \cite{Raine75,Padmanabhan82,Nikolic02}. As pointed out in \cite{Raine75} the difficulty of studying particle production in the presence of a non-asymptotically flat gravitational background, is that the definition of the particle production rate can depend on the method of calculation ({\it e.g.} using creation/annihilation operator versus using a definition of a vacuum state versus using a Feynman Green function).  In \cite{Raine75} the vacuum-to-vacuum amplitude was calculated in the path integral approach for Friedmann-Robertson-Walker space-time. The amount by which this amplitude differed from unity was used to obtain the particle production rate and it was found to give a different particle production rate from the usual method of diagonalization of the Hamiltonian. 
While this warning about different calculation methods yielding different particle production rates may not necessarily apply to the gravitational wave backgrounds, which are the focus of this paper \footnote{ This difference in particle production rates, calculated via different methods, occurs in space-times where it is not clear how to define asymptotic states. This is the case for the FRW space-time considered in \cite{Raine75} but is not the case for the gravitational wave background.}, we nevertheless mention it to point out the subtle issues which surround the definition of particles in curved space-times, and determining if a given space-time will have particle production associated with it or not. Finally we note that reference \cite{Nikolic02} gives arguments that for general {\it time-dependent} metrics the issue of the definition of particles and whether or not particle production occurs is still an unresolved problem.

In this paper, we obtain the particle production rate by calculating the conserved 4-current of a massless, scalar field in a gravitational wave background and then comparing this to the massless, scalar field in flat space-time. The difference between these two situations -- scalar field in a gravitational wave background versus scalar field in Minkowski space-time -- we take as a measure of the rate of particle production. This 4-current method for calculating the production rate and/or super radiance is similar to that used in references \cite{Wald,Nikolic02,Markus14,Stahl16}. The 4-current method employed here can also be compared to the work of Gertsenshtein \cite{Gertsenshtein60}, who investigated the production of electromagnetic radiation when a gravitational wave encountered a region of space-time with a uniform magnetic field. In Gertsenshtein the interaction between the magnetic field and gravitational wave background produced electromagnetic radiation. Here we replace the magnetic field by a massless scalar field. Recently, M{\"o}sta {\it et al.} \cite{mosta} carried out a study
similar to that of Gertsenshtein, where they studied the electromagnetic radiation produced by a gravitational waves coming from a
pair of inspiralling black holes embedded in a constant magnetic field.  

In section II we study the solution of a massless scalar field in a gravitational wave background and use this to calculate 
a particle production rate. In section III we use the results of section II to give a rough estimate an attenuation length for the gravitational wave due to the production of electromagnetic radiation from the gravitational wave. 

\section{Scalar field in gravitational wave background}

In this paper we use a massless scalar field as a stand-in for the more physically realistic massless photon. The justification 
for this is that one can write a vector field as $A_\mu (x_\nu) = \epsilon _\mu \varphi (x_\nu)$ where $\epsilon_\mu$ is the polarization
4-vector and $\varphi (x_\nu)$ is a scalar function which obeys the massless Klein-Gordon equation. The vector field $A_\mu$ certainly has
more degrees of freedom, because of $\epsilon _\mu$, as compared to a simple scalar field, but this would at most change the production rate
by some factor of order unity. At the level of individual quanta one can point to the work of reference \cite{Bohr14} where the cross sections for gravitons Compton scattering from scalar and vector particles is given {\it i.e.} the processes $S + g \rightarrow S + g$ and $\gamma + g \rightarrow \gamma + g$. These Compton scattering diagrams can be rotated to give graviton production processes {\it i.e. } $g + g \rightarrow S + S$ and $g + g \rightarrow \gamma + \gamma$. From \cite{Bohr14} the diagrammatic evaluation of both these two processes are non-zero, although they do differ by numerical factors of order unity due to the different spins of the scalar versus vector particles. 

With this justification we begin by placing a massless, scalar field in curved space-time and writing down the Klein-Gordon equation coupled to the space-time described by the metric $g_{\mu \nu}$,

\begin{equation}
\label{KGvacuum1}
\frac{1}{{\sqrt { -\left| {g_{\mu \nu } } \right|} }}\left( {\partial _\mu  g^{\mu \nu } \sqrt { - \left| {g_{\mu \nu } } \right|} \partial _\nu  } \right)\varphi  = 0.
\end{equation}

\noindent For our background we take a plane gravitational wave with the $+$ polarization. The metric for this can be 
written as \cite{Schutz},

\begin{equation}
ds^2 = -dt^2 + dz^2 + f(u)^2 dx^2 + g(u)^2 dy^2 ~.
\label{metric}
\end{equation}

\noindent The variables in the metric, $u = z-t$ and $v = z+t$, are light cone coordinates and the metric components $f(u)$ and $g(u)$ will be required to be oscillatory functions as expected for a gravitational wave background. The determinant of the metric is ${\left| {g_{\mu \nu } } \right|} = det[ g_{\mu \nu}$] and $\sqrt{-{\left| {g_{\mu \nu } } \right|}} = fg$. 
Substituting this in equation \eqref{KGvacuum1},

\begin{equation}
\frac{1}{{fg}} \left( - \partial _t (fg)\partial _t  + \frac{1}{{f^2 }}\partial _x (fg)\partial _x  + \frac{1}{{g^2 }}\partial _y (fg)\partial _y  + \partial _z (fg)\partial _z \right)\varphi  = 0.
\label{KGvacuum2}
\end{equation}

\noindent Since $u$ is only a function of $z$ and $t$ the expression can be expanded,

\begin{widetext}
\begin{equation}
\left( { - \partial _t^2  - \frac{1}{{fg}} {\partial _t (fg)} \partial _t  + \frac{1}{{f^2 }}\partial _x^2  + \frac{1}{{g^2 }}\partial _y^2  + \partial _z^2  + \frac{1}{{fg}} {\partial _z (fg)} \partial _z } \right)\varphi  = 0.
\label{KGvacuum3}
\end{equation}
\end{widetext}

\noindent Applying the chain rule for the $t$ and $z$ derivatives, ${\partial _t (fg) =  - \partial _u (fg)}$, ${\partial _z (fg) = \partial _u (fg)}$, $\left( {\partial _z^2  - \partial _t^2 } \right) = 4\partial _u \partial _v $, $\left( {\partial _t  + \partial _z } \right) = 2\partial _v $, $\left( {\partial _z  - \partial _t } \right) = 2\partial _u $,
and multiplying by $f^2g^2$,

\begin{equation}
\left( {4f^2 g^2 \partial _u \partial _v  + 2fg {\partial _u (fg)} \partial _v  + g^2 \partial _x^2  + f^2 \partial _y^2 } \right)\varphi  = 0.
\label{KGvacuum4}
\end{equation} 

\noindent At this point we are still looking at the exact solution to the Klein-Gordon equation using the metric of equation \eqref{metric}. To evaluate the solution for a weak gravitational wave the linearized gravity approximation will be introduced in the terms of the metric, $f\left( u \right) = 1 + \varepsilon \left( {ku} \right)$, and $g\left( u \right) = 1 - \varepsilon \left( {ku} \right)$ and substituted into equation \eqref{KGvacuum4}. Also note that the metric of equation \eqref{metric} describes a wave propagating in the $z$ direction and the $x$ and $y$ spatial directions must be physically indistinguishable. Based on the isotropy of space-time and assuming a non-thermal vacuum \cite{Modanese95,King12,Jones15}, we take $\left( {\partial _y^2 - \partial _x^2 } \right)\varphi=0$ as a property of our scalar field solution. Using this and collecting terms together equation \eqref{KGvacuum4} can be expressed as,

\begin{widetext}
\begin{equation}
\left[ {4\left( {1 - 2\varepsilon ^2  + \varepsilon ^4 } \right)\partial _u \partial _v  - 4\left( {1 - \varepsilon ^2 } \right)\varepsilon \left( {\partial _u \varepsilon } \right)\,\partial _v  + \left( {1 + \varepsilon ^2 } \right)\partial _x^2  + \left( {1 + \varepsilon ^2 } \right)\partial _y^2 } \right]\varphi  = 0.
\label{KGvacuum7}
\end{equation}
\end{widetext}

\noindent We now assume that $\varepsilon \left( ku \right) = h_ +  e^{i ku}$ {\it i.e.} oscillatory functions typical for 
gravitational waves in linearized general relativity. In this expression $h_{+}$ is the dimensionless amplitude of the gravitational wave. Substituting into equation \eqref{KGvacuum7} we obtain,

\begin{equation}
 \left( {4F\partial _u \partial _v  - 4ikG\,\partial _v  + H\left( {\partial _x^2  + \partial _y^2 } \right)} \right)\varphi  = 0,
\label{KGvacuum10}
\end{equation}

\noindent where,

\begin{equation}
\begin{array}{l}
 \quad F\left( {ku} \right)  \equiv  \left( {1 - 2h_ + ^2 e^{2iku}  + h_ + ^4 e^{4iku} } \right), \\ 
 \quad G\left( {ku} \right)  \equiv  \left( {h_ + ^2 e^{2iku}  - h_ + ^4 e^{4iku} } \right), \\ 
 \quad H\left( {ku} \right)  \equiv  \left( {1 + h_ + ^2 e^{2iku} } \right). \\ 
 \end{array}
\label{KGvacuum10B}
\end{equation}

\noindent Equation \eqref{KGvacuum10} is separable taking $\varphi  = X\left( x \right)Y\left( y \right) U \left( u \right) V \left( v \right)$ and identifying the eigenvalue equations for $X(x)$ and $Y(y)$ as,

\begin{equation}
\begin{array}{l}
 \partial _x^2 X =  - k^2 _x X \to X = e^{ik_x x},  \\ 
 \partial _y^2 Y =  - k^2 _y Y \to Y = e^{ik_y y}.  \\ 
 \end{array}
\label{XYequations}
\end{equation}
Note, that the $x$ and $y$ direction eigenfunctions are simply free waves as is to be expected since the gravitational wave is
in the $u=z-t$ direction. Setting $2k^2 _{xy}    \equiv   k_x^2  + k_y^2 $ and using \eqref{XYequations} we find that \eqref{KGvacuum10} becomes

\begin{equation}
F\frac{{\partial _u U}}{U}\frac{{\partial _v V}}{V} - ikG\frac{{\partial _v V}}{V} - H\frac{{k^2 _{xy} }}{2} = 0.
\label{ASequation}
\end{equation}

\noindent Now since the light front coordinate $v$ is orthogonal to $u$ and since the gravitational wave only depends
on $u$ one expects that the eigenfunction $V(v)$ also is solved by a free, plane wave, as was the case for
$X(x)$ and $Y(y)$. This is indeed the case and we find

\begin{equation}
 - i\partial _v V = k_v V \to V = e^{ik_v v} .
 \label{eigenvalueV}
\end{equation}

\noindent Substituting equation \eqref{eigenvalueV} into equation \eqref{ASequation} yields,

\begin{equation}
ik_v F\frac{{\partial _u U}}{U} + kk_v G - \frac{{k^2 _{xy} }}{2}H = 0.
 \label{eigenvalueU}
\end{equation}

\noindent Defining the eigenvalue $\lambda   \equiv   \frac{k_{xy}^2}{2k_v}$ \eqref{eigenvalueU} can be rearranged,

\begin{equation}
i\frac{{\partial _u U}}{U} = \lambda \frac{H}{F} - k\frac{G}{F}.
 \label{eigenvalueU3}
\end{equation}

\noindent This equation can be integrated to give,

\begin{equation}
U = A e^{\frac{\lambda }{k}} e^{  \frac{- \lambda }{{k\left( {1 - h_ + ^2 e^{2iku} } \right)}}} \left( {1 - h_ + ^2 e^{2iku} } \right)^{\frac{1}{2}\left( {\frac{\lambda }{k} - 1} \right)} e^{ - i\lambda u}  ~.
 \label{eigenvalueU4}
\end{equation} 

\noindent $A$ is a normalization constant which we will fix later. The first term ($e^{\frac{\lambda }{k}}$) is needed to insure that as 
$h_+ \to 0$ ({\it i.e.} the gravitational wave is turned off) that the eigenfunction for the $u$ direction becomes a free plane wave,
$e^{- i\lambda u}$. Collecting together all the terms in $x, y, v$ and $u$ directions gives the solution
of the scalar field in the gravitational wave background,

\begin{equation}
\varphi  = A e^{\frac{\lambda }{k}} e^{ - \frac{\lambda }{{k\left( {1 - h_ + ^2 e^{2iku} } \right)}}} \left( {1 - h_ + ^2 e^{2iku} } \right)^{\frac{1}{2}\left( {\frac{\lambda }{k} - 1} \right)} e^{ - i\lambda u} e^{ik_v v} e^{ik_x x} e^{ik_y y} .
\label{Sfield}
\end{equation}
This solution for the scalar field given in \eqref{Sfield} is very similar to solution found in \cite{Padmanabhan99} for the 
static electric field pair production evaluated in light front coordinates. Taking the limit  $h_+ \to 0$ of equation \eqref{Sfield} returns the expected Minkowski vacuum solution for the scalar field,

\begin{equation}
\varphi _0  = A e^{ - i\lambda u} e^{ik_v v} e^{ik_x x} e^{ik_y y} \rightarrow
 A e^{i\left( {k_v  + \lambda } \right)t} e^{i\left( {k_v  - \lambda } \right)z} e^{ik_x x} e^{ik_y y} .
 \label{ZeroSolution}
\end{equation}

\noindent In the last expression we have reverted from light front to Cartesian coordinates.
It is clear that the scalar field in \eqref{ZeroSolution} is a free wave. By defining an energy 
$k_{0}={k_v  + \frac{{k_{xy}^2 }}{{2k_v }}}$ and a momentum in the $z$-direction $k_{z}= k_v -{\frac{{k_{xy}^2 }}{{2k_v }} }$
and using the previously defined $k_x^2  + k_y^2  = 2k_{xy}^2$ one can check that energy-momentum of the free solution
in \eqref{ZeroSolution} satisfy the usual kinematic relationship for a free particle in Minkowski space-time namely 
$k_0^2  = k_x^2  + k_y^2  + k_z^2$. 

We now use the result for the scalar field given in \eqref{Sfield} to calculate the associated 4-current density which will
then allow us to calculate the rate of pair production of the scalar field from the gravitational wave background. 
The $u$ component of the scalar field 4-current is given in terms of $\varphi$ as

\begin{equation}
j_u  = - i\left( {\varphi ^* \partial_u \varphi   - \varphi \partial_u \varphi  ^* } \right) .
\label{current-u}
\end{equation}

\noindent Substituting $\varphi$ from \eqref{Sfield} into \eqref{current-u} we find that the time averaged 
$u$ component of the 4-current is, 

\begin{equation}
\langle j_u \rangle = - 2 A^ 2 \lambda  - 
A^2 \left( {\frac{9}{2}\frac{{\lambda ^3 }}{{k^2 }} - \frac{{12\lambda ^2 }}{k} + \frac{{13}}{2}\lambda  - k} \right)h_ + ^4 .
\label{current-u4two}
\end{equation}

\noindent The brackets represent the time averaging. In obtaining this expression we have taken the light front 
coordinate averages for the cosines,  $\left\langle {\cos ^2 \left( {2ku} \right)} \right\rangle  = \frac{1}{2}$, $\left\langle {\cos ^4 \left( {2ku} \right)} \right\rangle  = \frac{3}{8}$, and $\left\langle {\cos \left( {2ku} \right)} \right\rangle  = \left\langle {\cos \left( {4ku} \right)} \right\rangle  = 0$. Also we have dropped terms higher than $h_+^4$. 

We now examine various limits of Equation \eqref{current-u4two}. First, in the limit when the gravitational wave vanishes, $h_+ \rightarrow 0$, the current becomes $j_u \rightarrow -2 \lambda A^2 \rightarrow -1 / V$ where we have fixed the normalization 
constant $A = \frac{1}{\sqrt{V}} \frac{1}{\sqrt{2 \lambda}}$ by requiring that there be one particle per volume $V$. Another option for $A$ would be to use the condition that there be $2 \lambda$ particles per volume $V$ which would give $A = \frac{1}{\sqrt{V}}$. Section 4.3 of \cite{halzen} discusses the various normalization conditions for scalar fields. Second, in the presence of both the scalar field ($\lambda \ne 0$) and gravitational wave ($h_+ \ne 0$) equation \eqref{current-u4two} indicates how the current is modified by the potential represented by the gravitational background. For certain values of $\lambda$, $k$, and $h_+$ the current in \eqref{current-u4two} gives a larger outgoing current than incoming. This can be likened to the calculation of the Penrose super radiance process \cite{Wald} where one ``scatters" a real scalar field from a rotating black hole and the outgoing scalar field may have more energy than the incoming field. Finally, one can take the limit $\lambda \to 0$, $k_v \to 0$ and $k_{xy} \to 0$ {\it i.e.} the initial scalar field is taken to its vacuum state. In this way one obtains what is called the Minkowski persistence amplitude \cite{Dunne09}. Because of the definition $\lambda \equiv \frac{k_{xy}^2}{2k_v}$ the limit $\lambda \to 0$ also means $k_{xy} \to 0$. In this limit the scalar field and its 4-current \eqref{Sfield} {\it do not} reduce to the vacuum case ({\it i.e.} $\varphi_0 \to 0$ and $j_u \to 0$) but rather reduce to
 
\begin{equation}
\varphi  \to \frac{1}{\sqrt {V}} \frac{1}{\sqrt{2 k}}\left( {1 - h_ + ^2 e^{2iku} } \right)^{-\frac{1}{2}} ~~~~~{\rm and} 
~~~~ j_u \to \frac{1}{V} h_+^4~.
\label{production}
\end{equation}
In \eqref{production} we have written out explicitly the normalization constant $A = \frac{1}{\sqrt {V}} \frac{1}{\sqrt{2 k}}$.
As before $V$ is the volume in which the scalar field is placed. This normalization of $\varphi$ (especially the $\frac{1}{\sqrt{2 k}}$
factor) is consistent with the normalization found in \cite{Stahl16} via the Wronskian condition. Note that $j_u$ in \eqref{production} is of order $h_+ ^4$. There are $h_+ ^2 e^{\pm 2iku}$ terms in $j_u$ that arise when one substitutes $\varphi$ from \eqref{production} into \eqref{current-u}. However these terms time average to zero. In \eqref{current-u} it is only terms that involve products of things like $h_+ ^2 e^{2iku}$ and  $h_+ ^2 e^{-2iku}$, which give a non-zero value after time averaging. This explains why one is justified in keeping the metric and $\varphi$ to order $h_+^2$ while the current derived from $\varphi$ is of order $h_+^4$. If one could write out the metric to order $h_+^4$ then $\varphi$ should have terms like $h_ + ^4 e^{4iku}$. However these terms, when run through the definition of the current in \eqref{current-u}, would time average to zero unless they were products of terms like $h_+ ^4 e^{4iku}$ and $h_+ ^4 e^{-4iku}$. 
These ``direct" product terms would be of order $h_+^8$ and would contribute to the current $j_u$, but the highest non-zero
terms would still be of order $h_+ ^4$ since the $h_+ ^2$ terms time average to zero. Terms from \eqref{current-u}
which were ``cross" products of things like $h_+ ^4 e^{4iku}$ and $h_+ ^2 e^{-2iku}$ or $h_+ ^4 e^{4iku}$ and $1$ time average to 
zero. There are other cases in general relativity where the metric is of lower order in $h_+$ as compared to the quantity calculated from
the metric. One common example is the energy carried by a linearized gravitational wave where the metric is kept to order $h_+$ while the energy-momentum calculated from this metric is to order $h_+ ^2$ \cite{cheng}. As a final comment on the order of $h_+$ we note that
if one only kept terms of order $h_+^2$ in the scalar field equation \eqref{KGvacuum7} then the vacuum limit for the scalar field 
given in \eqref{production} would become $\varphi  \propto \left( {1 - 2 h_ + ^2 e^{2iku} } \right)^{-\frac{1}{4}}$ which to order
$h_+ ^2$ has the same expansion as $\varphi$ from \eqref{production} and yields the same $j_u$ as in \eqref{production} to
order $h_+^4$ 

The result in \eqref{production} can be related to the Higgs mechanism \cite{Higgs} where a scalar field develops a 
non-zero vacuum expectation value of $\varphi = \sqrt{\frac{m^2}{2 \lambda}}$ due to a potential with a quartic {\it self interaction} term plus tachyonic mass term ({\it i.e.} $-m^2 \varphi ^2 + \lambda \varphi ^4$). The self interacting scalar potential in the usual Higgs mechanism is time independent. In the present case the scalar field develops a non-zero vacuum expectation value (the scalar field expressison in \eqref{production}) due to the background gravitational wave potential. Because of the space and time dependent nature of the background gravitational field the vacuum expectation value from \eqref{production} is also space and time dependent -- the $e^{2iku}$ term in the expression for $\varphi$. Since the vacuum expectation value in this case is space and time dependent, one has a non-zero 
4-current in the $u=z-t$ direction, $j_u =  \frac{1}{2V} h_+^4$, as opposed to the usual Higgs mechanism case where the constant vacuum expectation value of the scalar field gives a zero 4-current associated with $\varphi = \sqrt{\frac{m^2}{2 \lambda}}$. Another difference between the present example and the canonical Higgs mechanism, is that in the present case the interaction that leads to the vacuum expectation value of $\varphi$ in \eqref{production} comes from the interaction between the scalar field and the gravitational field. In the canonical Higgs mechanism the vacuum expectation value is due to the $\lambda \phi^4$ self interaction of the scalar field. Thus the non-zero scalar vacuum expectation value of the present case can be compared to the version of the Higgs mechanism that occurs in superconductors, where it is the phonons of the background lattice that are responsible for the interaction that binds electrons into Cooper pairs and which leads to superconductivity.  

The important point about \eqref{production} is that $j_u \ne 0$ even though we have taken the scalar field to its Minkowski vacuum state. We interpret this non-zero $j_u$ as being connected to a non-zero production rate of the scalar field by the gravitational wave background. (In the next section we draw the exact connection between $j_u$ and the production rate). That one should get a non-zero result for the process of gravitons converting to these scalar ``photons" is supported by the Feynman diagram amplitudes like $g + g \to \gamma + \gamma$ which are non-zero \cite{Skobelev75, Bohr14, Bohr15}.              

The calculation of the production of the scalar field via the time varying gravitational wave background of \eqref{metric} can be compared to the similar calculation for de Sitter space-time from reference \cite{Akhemedov09}. There a massive scalar field was placed in the time-dependent de Sitter space-time and the amplitude of the scalar field in the de Sitter background was used to determine the scalar field production rate at the expense of the gravitational field. Unlike the de Sitter space-time metric there is no horizon in the gravitational wave metric.

In the above discussion the ansatz function, $f(u)$ , $g(u)$ were not exact solutions to the plane wave space-time
of \eqref{metric}. We now briefly show that one obtains similar results for an exact plane wave, oscillatory metric, showing that the results are not an artifact of the approximate metric. 

In order for $f(u)$ and $g(u)$ in \eqref{metric} to be exact solutions to the Einstein field equations they  
need to satisfy the condition ${\ddot f}/ f + {\ddot g}/ g = 0$ \cite{Schutz}. A simple exact, plane wave, solution is 
given by $f = e^{iku} e^{ - ku}$ and $g = e^{iku} e^{ku} $. These ansatz functions have oscillatory wave parts ($e^{iku}$) but they also have exponentially growing or decaying amplitudes ($e^{ \pm ku}$). Near $u=0$ one has oscillating, wave solutions due to the $e^{iku}$ parts of the ansatz function, but as $u$ moves away from $u=0$ the $e^{\pm ku}$ terms act like growing/decaying  amplitudes. Because of this these solutions can only be of use for a restricted range of $u$ near $u=0$. Asymptotically, as $u\rightarrow \infty$, the functions $f(u), g(u)$ are not physically acceptable. Substituting $f = e^{iku} e^{ - ku}$ and $g = e^{iku} e^{ku}$ into equation \eqref{KGvacuum4},

\begin{widetext}
\begin{equation}
\left( {4 e^{4iku} \partial _u \partial _v  + 2 e^{2iku} \partial _u \left( {e^{2iku} } \right)\partial _v  + e^{2iku} e^{2ku} \partial _x^2  + e^{2iku} e^{ - 2ku} \partial _y^2 } \right)\varphi  = 0,
\label{Exact1}
\end{equation}
\end{widetext}

\noindent and making the substitution $\varphi  = U(u) V(v) X(x) Y(y) = U(u) e^{ik_v v} e^{ik_x x} e^{ik_y y}$,

\begin{equation}
\left( {i \frac{{\partial _u U}}{U} -  k - e^{ - 2iku} e^{2ku} \frac{{k_x^2 }}{{4k_v }} - e^{ - 2iku} e^{ - 2ku} \frac{{k_y^2 }}{{4k_v }}} \right) = 0.
\label{Exact2}
\end{equation}

\noindent In the limit when the gravitational wave is absent ({\it i.e.} $k \to 0$) the solution to \eqref{Exact2} is again given by \eqref{ZeroSolution}. When $k \ne 0$ the solution is \eqref{Exact2},

\begin{equation}
U = A e^{\left( {\frac{{\left( {1 - i} \right)}}{{4k }}\lambda _x e^{ - 2iku} e^{2ku}  + \frac{{\left( {1 + i} \right)}}{{4k}}\lambda _y e^{ - 2iku} e^{ - 2ku} } \right)} e^{-iku} ,
\label{Exact3}
\end{equation}

\noindent As before $A$ is a constant and $\lambda _x  \equiv \frac{{k_x^2 }}{{4k_v }}$ and $\lambda_y \equiv \frac{{k_y^2 }}{{4k_v }}$. As before if we take the limit of the massless scalar field to its vacuum state  ({\it i.e.} taking the limit $k _x  \to 0$ , $k _y  \to 0$ , $\lambda _{x,y} \to 0$ and $k_v \to 0$) one finds $U (u)  \to \frac{1}{\sqrt{V}} \frac{1}{\sqrt{2 k}}e^{-iku}$ so that as before $\varphi$ does not go to zero but rather 
$\varphi \to \frac{1}{\sqrt{V}} \frac{1}{\sqrt{2k}} e^{-iku}$. In this limit we have again written out the normalization constant 
as $\frac{1}{\sqrt{V}} \frac{1}{\sqrt{2k}}$. As before we can calculate the current in the $u$ direction in this limit and find that,

\begin{equation}
j _u= \mathop {\lim }\limits_{\left( {k_x ,k_y } \right) \to 0} - i \left( {U^* \partial _u U  - U \partial _u U^* } \right) = \frac{1}{V} ~.
\label{Exact3a}
\end{equation}

\noindent There is no explicit amplitude, $h_+$, in this case since the changing amplitudes of the ansatz functions, $f(u) , g(u)$, are given 
by $e^{\pm ku}$. 

There are other exact plane wave solutions one could examine. The simplest is $f(u)=g(u) =u$ \cite{bondi} which represents
a plane wave pulse rather than an oscillatory wave. Performing the above analysis with $f(u)=g(u) =u$ leads to $j_u =0$ in the vacuum
limit rather than the non-zero result of \eqref{production} or \eqref{Exact3a}. Thus the non-zero result for $j_u$ for the oscillatory 
ansatz functions is non-trivial.

\section{Estimated attenuation length}

In this section we use the vacuum current $j_u$ of the last section to estimate the production rate of massless quanta from
the gravitational wave background. Other, recent works which connect the particle production rate with currents in curved space-times, can be found in \cite{Nikolic02,Markus14,Stahl16}. In \cite{Markus14} the connection between the current and the production rate per unit volume is given by
\begin{equation}
\label{j-gamma}
\frac{\Gamma}{V} \Delta T \approx | j_u | ~,
\end{equation}
where $\Delta T$ is a characteristic time for the problem and $V$ is the volume of the system as before. Using $|j_u| = \frac{h_+ ^4}{V}$ 
from \eqref{production} and taking the characteristic time as $\Delta T \sim \frac{1}{\omega}$, where $\omega$ is the frequency
of the gravitational wave, we arrive at the production rate
\begin{equation}
\Gamma   =  \omega h_ + ^4 ~.
\label{TRate}
\end{equation}

We now use this production rate to estimate the effect this has on the decay of the amplitude $h_+$.  
We ignore the effect of the usual $\frac{1}{r}$ fall off in $h_+$  due to spherical nature of the outgoing gravitational wave. If
we take $N_g$ as the number of gravitons, then the standard result for the change of $N_g$ due to $\Gamma$ reads
\begin{equation}
\label{ex-decay}
\frac{dN_g}{dt} = - \Gamma N_g \to c \frac{dN_g}{dz} = - \Gamma N_g ~.
\end{equation}
In anticipation that we will be more interested in a decay length than a decay time we have taken
$dt \rightarrow dz/c$. As a starting assumption we will take the number of gravitons as $N_g \propto h_+^2$, which is motivated by a similar relationship in QED where the number of photons is related to the square of the vector potential, $N_\gamma \propto A_\mu A^\mu$. In this way, and using the decay rate from \eqref{TRate}, equation \eqref{ex-decay} becomes
\begin{equation}
\label{ex-decay-1}
\frac{d (h_+^2)}{dz} = - c \omega h_+ ^4 (h_+ ^2 ) \to \frac{d h_+}{dz} = - \frac{1}{2} k h_+ ^5 ~,
\end{equation}   
where $k= c \omega$. Equation \eqref{ex-decay-1} has the solution
\begin{equation}
\label{ex-decay-2}
h_+ (z) =\left( 2 k z + K_0 \right) ^{-1/4} ~, 
\end{equation}   
where $K_0 = \left( h_+ ^{(0)} \right)^{-4}$ and $h_+ ^{(0)}$ is the value of $h_+$ at $z=0$. What \eqref{ex-decay-2} shows is that for large distances ({\it i.e.} large $z$) that $h_+$ falls off like $\propto z^{-1/4}$ which is slower than the $z^{-1} \sim r^{-1}$ fall off due to the spherical nature of the outgoing gravitational wave. Thus the main factor in determining the fall off of $h_+$ at {\it large distances} is just the usual $\frac{1}{r}$ fall off. However near the source of the gravitational wave, $z=0$, the fall off in $h_+$ due to the conversion of the gravitational wave field into the massless field could be important. 
We can use \eqref{ex-decay-2} to make an estimate of the attenuation length, $\Lambda$, of the gravitational wave due to the conversion 
into electromagnetic radiation. If we define the decay length $\Lambda$ as the distance over which $h_+ (z=\Lambda) = \frac{1}{2} h_+ ^{(0)}$
we find that 
\begin{equation}
\label{lambda}
\Lambda = \frac{15}{2 k \left( h_+ ^{(0)} \right)^{4}} \to \frac{1.5 \times 10^7 }{2 \left( h_+ ^{(0)} \right)^{4}} ~ {\rm meters}~.
\end{equation}
In the last expression we have assumed a frequency of $\omega \sim 300$ Hz which yields $k =\frac{\omega}{c} \sim 10^{-6} ~ m^{-1}$. 
This estimate of the frequency is a very simple and rough estimate based on the upper range of the ``chirp" for GW150914.
In Table I we give different $\Lambda$'s for different initial gravitational wave amplitudes $h_+^{(0)}$.

\vskip 0.3cm
\begin{table}[!ht]
\centering
\begin{tabular}{|c|c|}
\hline  ~~$h_+^{(0)}$ ~~  & \ $ ~~\Lambda ~ (m) ~~  $     
\\   
\hline  $10^{-21}$ & \ $10^{89}  $          
\\  
\hline  $10^{-15}$ & \ $10^{65} $   
\\
\hline  $10^{-9}$ & $10^{41} $  
\\  
\hline  $10^{-5}$ & $10^{25} $  
\\  
\hline $10^{-3}$ & \ $10^{17} $  
\\  
\hline
\end{tabular}
\caption{Various values of the decay length $\Lambda$ versus $h_+ ^{(0)}$ for $\omega \approx 3 \times 10^{2} Hz$ and
$k \approx 10^{-6} ~ m^{-1}$. We begin with $h_+ \approx 10^{-21}$ which is roughly the measured strain reported for GW150914 \cite{LIGO}.}
\end{table}

The size of the observable Universe is approximately $10 ^{27}$ meters so from Table I we see that the estimated decay length will be important only if $h_+ ^{(0)}$ is fairly large -- of the order $10^{-5}$ or larger.  The take away message from Table I is that the conversion of the gravitational background wave into the massless field is significant only close to the source of the gravitational waves where $h_+$ is large.  

\section{Discussion and Conclusions}

In this paper we have looked at the possibility that a gravitational wave background could create massless scalar particles/fields. The massless
scalar field quanta where taken as a simplified model of a photon. This is similar to the Schwinger effect but with the static electric field replaced by a gravitational wave background and the electron/positron replaced by massless scalar ``photons". Since the created field was massless, we did not have the exponential suppression of the particle production rate which one finds in the Schwinger effect.  Because of this lack of exponential suppression one expects the conversion of gravitaitonal wave into electromagnetic radiation to potentially play a more prominent, physical role. In particular we suggested that the creation of photons at the expense of the gravitational wave field would lead to an additional fall off of the dimensionless amplitude $h_+$ with distance from the source, in addition to the usual $\frac{1}{r}$ fall off. Based on the production rate per unit volume given in \eqref{TRate} we made an estimate of the decay length for various amplitudes $h_+$, given in Table I. Unless $h_+ > 10^{-5}$ our estimate for the decay length, $\Lambda$, given in Table I, was so large that one would not expect this process to weaken the gravitational wave even over cosmological distances. This was in agreement with the conclusions based on Feynman diagram calculations \cite{Skobelev75}. But close enough to the source one will have $h_+ \ge 10^{-5}$ so in this {\it near} region one might expect the attenuation to be important. Since we used a gravitational plane wave this ignored the $\frac{1}{r}$ fall off for a more realistic spherical wave. The overall conclusion, both from Table I and from \eqref{ex-decay-2}, was that the production of the massless field, $\varphi$, coming from the gravitational wave background and the subsequent attenuation of the gravitational wave background would only be important near the source of the gravitational wave. 

One of the main results of this paper were the calculation of the scalar field \eqref{Sfield} and associated 4-current \eqref{current-u4two}
in the case when the scalar field is placed in a gravitational wave background. In the vacuum limit 
({\it i.e.} $k _x  \to 0$ , $k _y  \to 0$ , $\lambda _{x,y} \to 0$ and $k_v \to 0$) the scalar field and 4-current took the non-zero limits $\varphi  \to \frac{1}{\sqrt{V}} \frac{1}{\sqrt{2k}} \left( {1 - h_ + ^2 e^{2iku} } \right)^{-\frac{1}{2}}$ and $j_u \to \frac{1}{V} h_+^4$. This can be likened to a time-dependent, Higgs-like mechanism where the scalar field develops a non-zero vacuum expectation value. The difference from the usual Higgs mechanism is that, here the effect is driven by the interaction of the scalar field with a gravitational background instead of with a self interaction $({\it i.e.}  \lambda \phi^4 )$. 
Also here the vacuum value of the scalar field is space-time dependent. This Higgs-like mechanism via the gravitational background
can be compared to the symmetry breaking that occurs in superconductors, where it is the background lattice and phonons which 
provide the mechanism leading to a non-zero expectation value for Cooper pairs. This connection to the Higgs mechanism will be discussed further in an upcoming paper \cite{Jones17}. 

Finally there are two predictions of physical phenomenon  that would occur if the production of photons from the gravitational wave 
background were significant. First, the amplitude $h_+$ measured by a detector on Earth would be smaller due to the fact that this amplitude would decrease not only from the $\frac{1}{r}$ fall off for an outward traveling wave, but also the amplitude would decrease as $r^{-1/4}$ due to the production of photons from the gravitational wave background. From the $r^{-1/4}$ dependence of the particle production rate one can see that this effect would only be important relatively close to the source. Second, the gravitational wave would produce electromagnetic radiation/photons traveling in the same direction as the initial gravitational wave. In fact it has already been suggested \cite{loeb} that a $\gamma$-ray signal which was detected in the same time frame as the gravitational wave signal, might be related to the detected gravitational wave. If the production mechanism of electromagnetic radiation from the gravitational wave background proposed here occurs and is significant, then we would predict that the gravitational wave signal should also be accompanied by an electromagnetic signal. 
However, in our process this electromagnetic signal should have roughly the same frequency as that of the gravitational wave. Thus we would predict that the electromagnetic wave coming from the gravitational wave would have extremely long wavelengths, on the order of 100s of kilometers {\it i.e.} the associated electromagnetic wave would have very large wavelengths. These wavelengths are of such a length that they could easily have gone undetected up to now.         

\section*{Acknowledgment}

PJ would like to thank Darrel Smith for making the authors aware of the research conducted by Michael Faraday \cite{Faraday1885} on the relation between gravity and electricity. 
DS is supported by grant $\Phi.0755$ in fundamental research in Natural Sciences by the Ministry of Education and Science of Kazakhstan.

\end{document}